\documentclass[11pt,twoside]{article}
\usepackage{asp2010}
\usepackage{natbib}                

\resetcounters

\begin{document}

\title{Fundamental Properties of Cool Stars with Interferometry}
\author{T. S. Boyajian$^{1,2}$, K. von Braun$^3$, G. van Belle$^4$, T. ten Brummelaar$^1$, D. Ciardi$^3$, T. Henry $^1$, M. Lopez-Morales$^{2,5,6}$, H. McAlister$^1$, S. Ridgway$^7$, C. Farrington$^1$, P. J. Goldfinger$^1$, L. Sturmann$^1$, J. Sturmann$^1$, N. Turner$^1$}

\affil{$^1$Center for High Angular Resolution Astronomy and Department of Physics and Astronomy, Georgia State University, P. O. Box 4106, Atlanta, GA 30302-4106, USA}
\affil{$^2$Hubble Fellow.}
\affil{$^3$NASA Exoplanet Science Institute, California Institute of Technology, MC 100-22, Pasadena, CA 91125, USA}
\affil{$^4$European Southern Observatory, Karl-Schwarzschild-Str. 2, 85748 Garching, Germany}
\affil{$^5$Department of Terrestrial Magnetism, Carnegie Institution of Washington, 5241 Broad Branch Road, NW, Washington, DC 20015, USA}
\affil{$^6$Institut de Ci\`{e}ncies de L'Espai (CSIC-IEEC), Spain}
\affil{$^7$National Optical Astronomy Observatory, P.O. Box 26732, Tucson, AZ 85726-6732, USA}

\begin{abstract}
We present measurements of fundamental astrophysical properties of nearby, low-mass, K- and M-dwarfs from our DISCOS survey (DIameterS of COol Stars). The principal goal of our study is the determination of linear radii and effective temperatures for these stars. We calculate their radii from angular diameter measurements using the CHARA Array and Hipparcos distances. Combined with bolometric flux measurements based on literature photometry, we use our angular diameter results to calculate their effective surface temperatures.  We present preliminary results established on an assortment of empirical relations to the stellar effective temperature and radius that are based upon these measurements. We elaborate on the discrepancy seen between theoretical and observed stellar radii, previously claimed to be related to stellar activity and/or metallicity. Our preliminary conclusion, however, is that convection plays a larger role in the determination of radii of these late-type stars. Understanding the source of the radius disagreement is likely to impact other areas of study for low-mass stars, such as the detection and characterization of extrasolar planets in the habitable zones.
\end{abstract}

\section{Introduction}		

Advances in high-resolution astronomical techniques, particularly in long-baseline optical interferometry (LBOI), have enabled us to resolve directly the disks of nearby stars.  Interferometric observations of nearby, low-mass, stars provide us with direct stellar size and effective temperature measurements.  These measurements provide a crucial resource in constraining stellar model atmospheres and stellar evolutionary models.

For KM dwarfs, in particular early M, the onset of atmospheric convection is a phenomenon that is often imperfectly addressed in such models. As seen in the results of \citet{ber06}, notable disagreements exist between interferometrically determined radii for M dwarfs and the ones calculated in low-mass stellar models in the sense that interferometrically obtained values for the stellar diameters are systematically larger, by more than 10\%, than those predicted from models. These results confirm other observations of larger-than-expected stellar radii, such as those of \citet{lop05}, \citet{von08} and \citet{boy08}, and provide motivation for adjustments to be made to models in order to match the observations
. 
\subsection{Observations}		

\begin{figure}[!ht]
\plotone{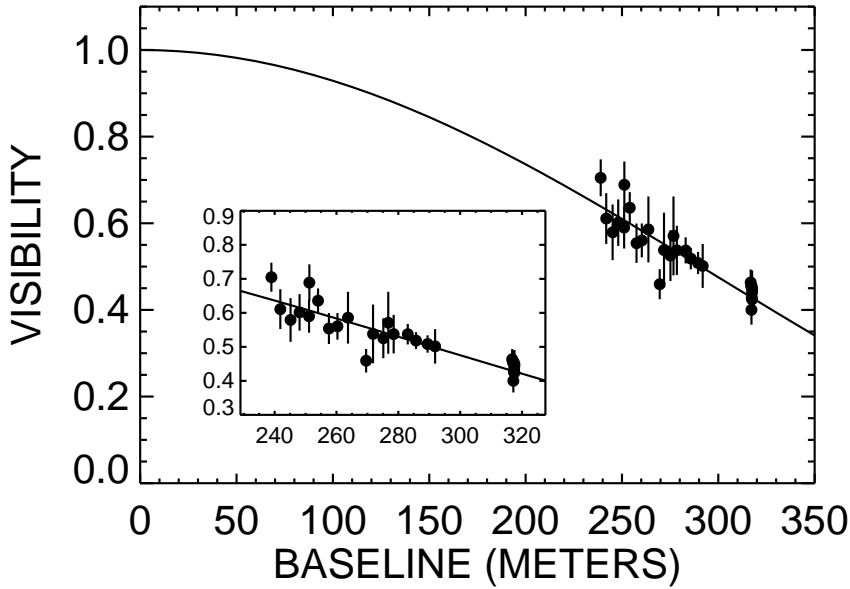}
\caption{Example limb-darkened angular diameter fit to CHARA visibilities measured in $K^{\prime}$-band for GJ~892.}
\label{fig:diameter}
\end{figure}

In this work, we present observations made at the CHARA Array, an optical/infrared interferometer located at Mount Wilson Observatory, California.  Our sample selection consists of $\sim$K0$-$M4 dwarfs, complete out to $\sim6.5$~parsecs and is limited only upon $V$-band magnitude and declination $>10$~degrees. Our goal is to measure the angular diameters of these stars to better than 4\%. Figure~\ref{fig:diameter} demonstrates our data fit to a limb-darkened diameter with 1\% precision to the fit. 

Thus far, we have measured the diameters of 23 cool dwarfs, 9 K-stars and 14 M-stars, to an average precision of $\sim$1.5\%. This more than doubles the current count of low mass stars with published diameters in the literature measured with interferometry to better than 5\% precision (see Table~\ref{tab:status})\footnote{A total of 17 diameters that meet this criteria, which includes 8 K-stars and 9-M-stars \citep{lan01, seg03, dif04, ber06, boy08, ker08a, dem09}.}. 

\begin{table}[!ht]
\caption{Status of Stellar Angular Diameters Measured to Better than 5\% Precision}
\label{tab:status}
\smallskip
\begin{center}
{\small
\begin{tabular}{ccc}
\tableline
\noalign{\smallskip}
 & Published & THIS WORK \\
\noalign{\smallskip}
\tableline
\noalign{\smallskip}
K-stars	&	8	&	9	\\
M-stars	&	9	&	14	\\
\noalign{\smallskip}
\tableline
\noalign{\smallskip}
TOTAL	&	17	&	23	\\
\noalign{\smallskip}
\tableline
\end{tabular}
}
\end{center}
\end{table}

\subsection{Fundamental Properties}				

\begin{figure}[!ht]
\plotone{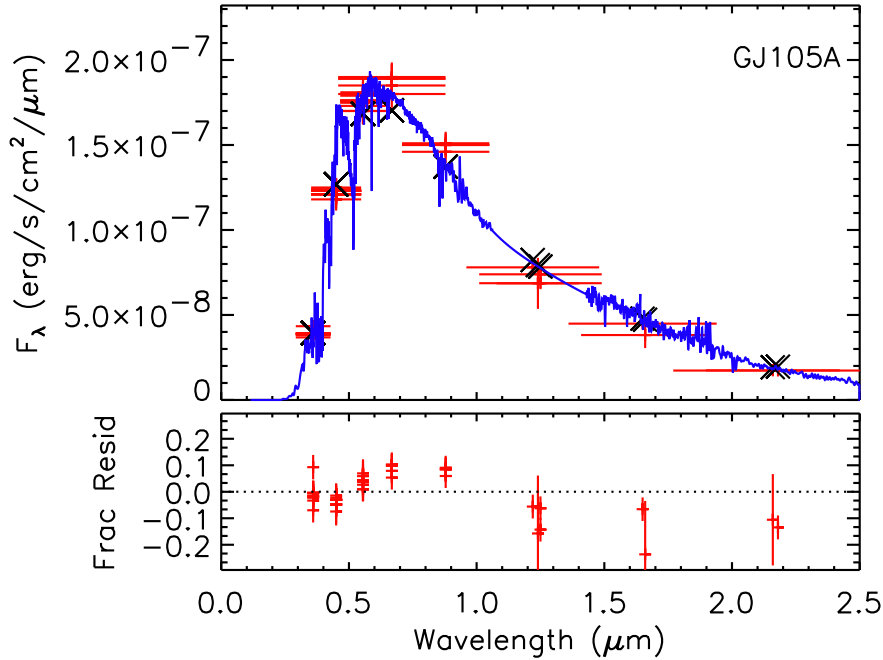}
\caption{Example SED fit.  {\it TOP:} The red markers are the input photometry with corresponding flux errors (y-direction) and bandwidth dimensions (x-direction). The black crosses indicate the predicted model photometry based upon the source flux distribution (blue line) and transmission of the filter at that bandpass.  {\it BOTTOM:} The fractional residuals to the above fit.  Flux errors are correctly displayed, but bandwidth dimensions are excluded in this plot for clarity.}
\label{fig:sedfit}
\end{figure}

The linear radii $R$ of these stars are quickly determined by combining the {\it HIPPARCOS} parallax with the interferometric angular diameter measurement. Furthermore, for each target we are able to calculate the bolometric flux $F_{\rm BOL}$ by performing a Spectral Energy Distribution (SED) fit to all available photometry in the literature (see Figure~\ref{fig:sedfit}), and again with {\it HIPPARCOS} parallax this quantity is readily converted to a measure of the absolute stellar luminosity $L$.  Finally, the effective temperature $T_{\rm EFF}$ can be expressed in relation to observable quantities using the Stephan-Boltzmann equation in the form $T_{\rm EFF} = 2341 (F_{\rm BOL} / \theta_{\rm LD}^2)^{0.25}$, where $F_{\rm BOL}$ is expressed in units of $10^{-8}$~erg/s/cm$^{2}$ and $\theta_{\rm LD}$ is in units of milli-arcseconds.  Generally speaking, the errors in radii and $T_{\rm EFF}$ are dominated by the errors in the measured angular diameter\footnote{The errors in $F_{\rm BOL}$ and $\pi_{\rm HIP}$ are negligible and/or non-influential in the overall error budget.}, where $\sigma_{\theta_{\rm LD}} \propto \sigma_{R} \propto \frac{1}{2}\sigma_{T_{\rm EFF}}$.

\section{Discussion}		
	
	\subsection{Empirical Relations}	

The inclusion of our data, which more than doubles the sample size presently available in the literature (Table~\ref{tab:status}), supplies us with a hearty foundation for establishing empirically calibrated relationships to the stellar radius and temperature.  This is especially the case when we constrain the precision of the data set to only those stars possessing have accurately measured properties, and in the following analysis, we choose a threshhold of $\sigma\theta_{\rm LD} < 5$\%.

\begin{figure}[!ht]
\plottwo{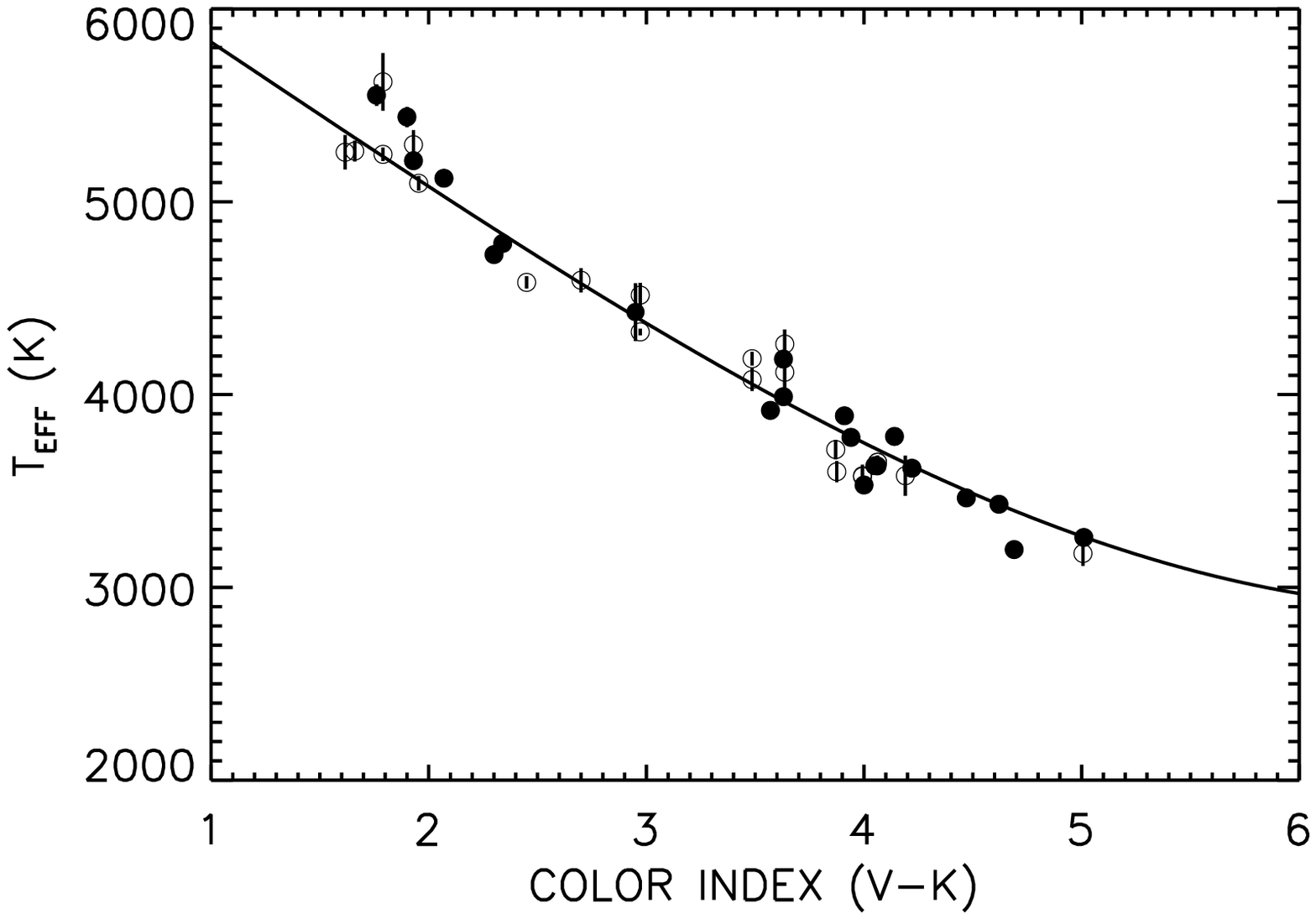}{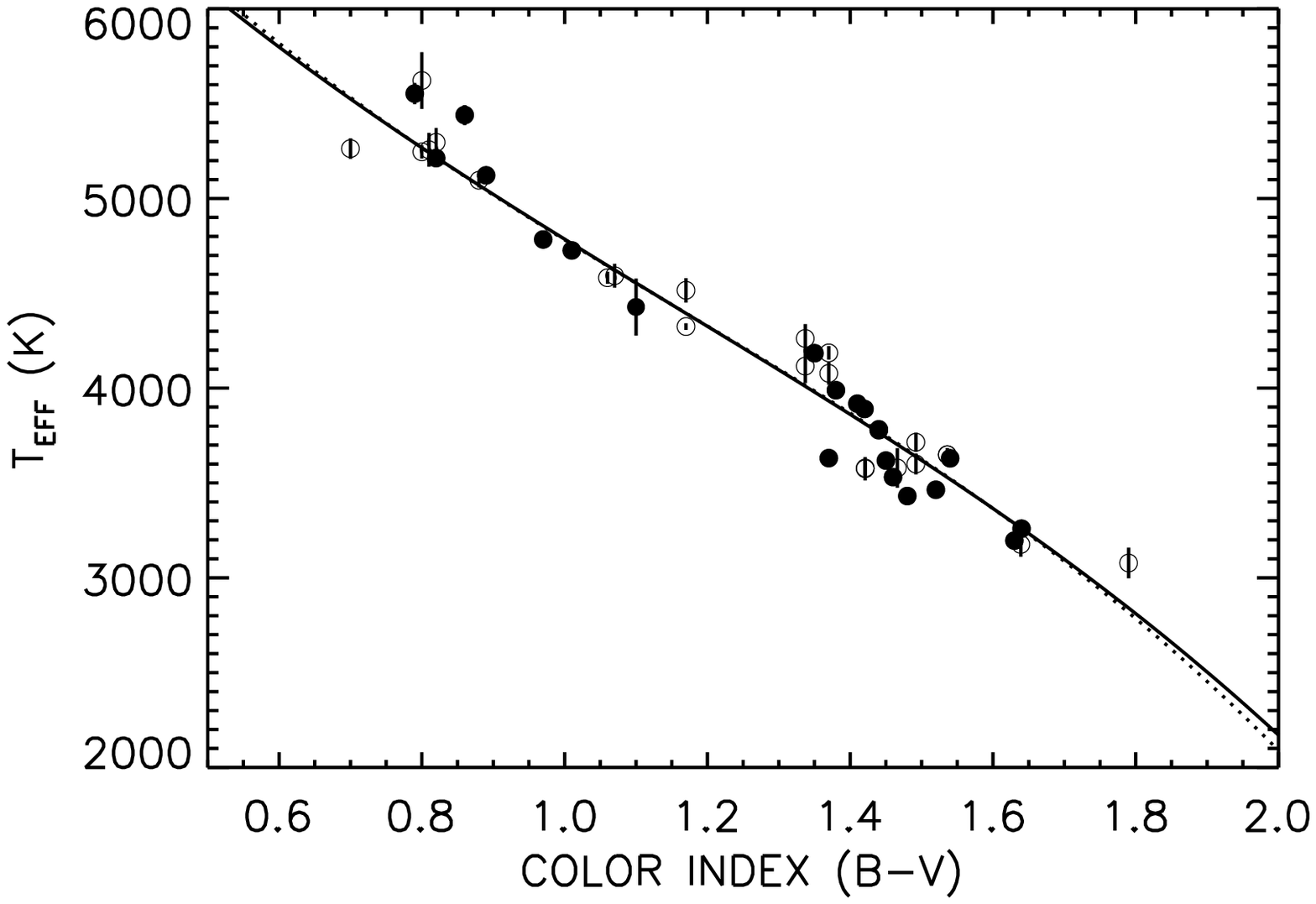}
\caption{Effective temperature versus ($V-K$) color index (left) and ($B-V$) color index (right).  Filled points are new results presented here, where open points are published values in the literature.  The 1$\sigma$ errors in $T_{\rm EFF}$ are shown, but in the majority of cases it tends to be smaller than the data point. A preliminary result to a $3^{rd}$-order polynomial is shown as a solid line.}
\label{fig:teff_VS_VmK_and_BmV}
\end{figure}

First, we present relations for the effective temperature of the star. In Figure~\ref{fig:teff_VS_VmK_and_BmV}, we show a preliminary fit to a $3^{rd}$-order polynomial with the ($V-K$) and ($B-V$) color indicies, both of which give a median absolute deviation from the fit of just over 100~$K$.  However, since the median error of the temperature measurements is a little more than half of these values, improvements to the fit are warranted.  An obvious improvement would be in the quality of the infrared photometry\footnote{The infrared colors for most these stars are saturated in {\it 2MASS}.  For example, the {\it 2MASS} $K$-magnitude for GJ~411 is $K = 3.501 \pm 0.352$)}.  We expect, but cannot confirm, an improvement with the $T_{\rm EFF}$ versus ($V-K$) relation from the $T_{\rm EFF}$ versus ($B-V$) relation since the ($V-K$) color index is less sensitive to metallicity compared to the ($B-V$) color index.  Additionally, in theory, a multi-parameter Temperature:Color:Metallicity fit would also improve the relation.  However, metallicities of these stars are very difficult to measure, and although a lot of work is being done with the M-type star metallicity calibrations, the K-dwarfs are currently being ignored.  

\begin{figure}[!ht]
\plottwo{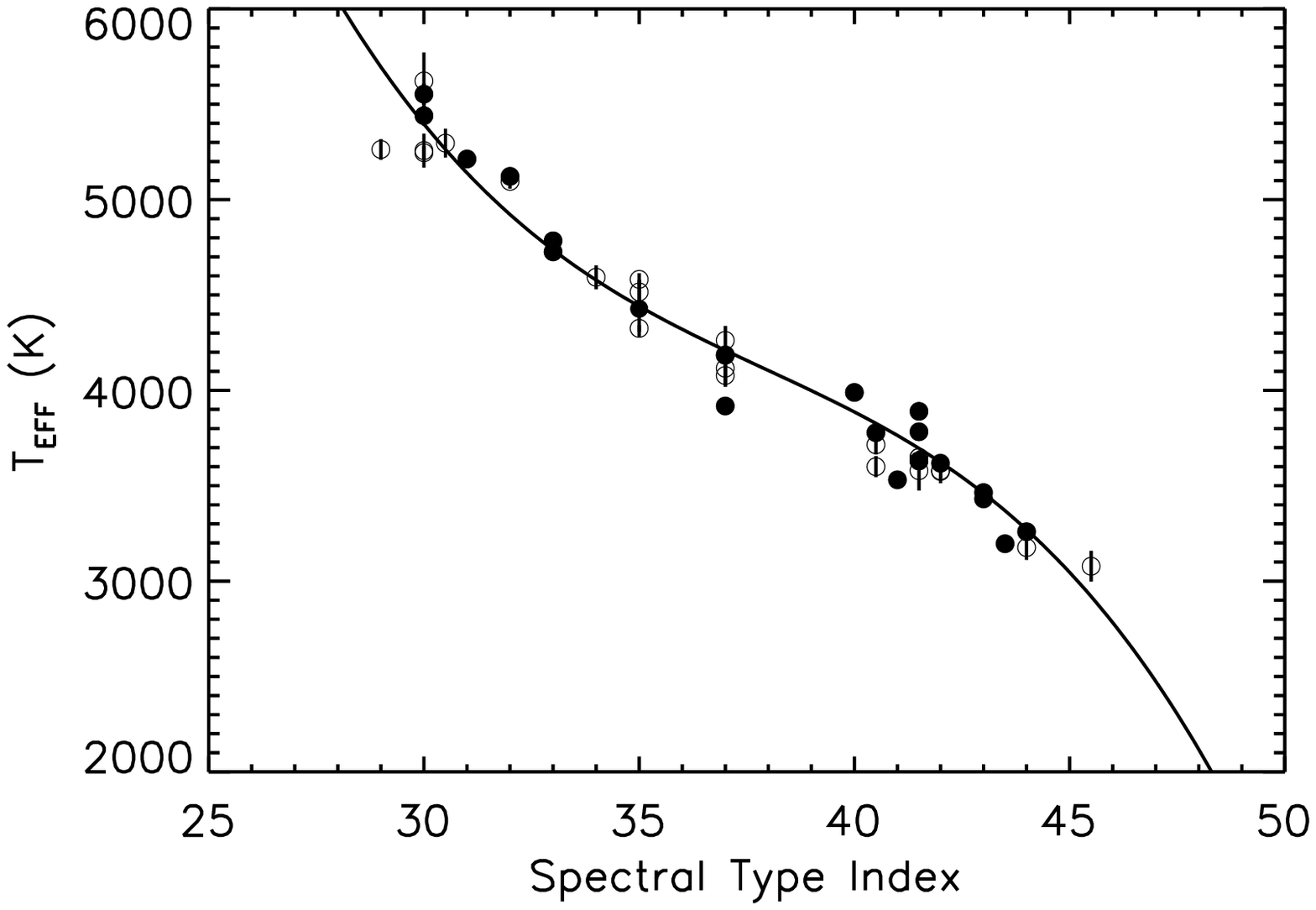}{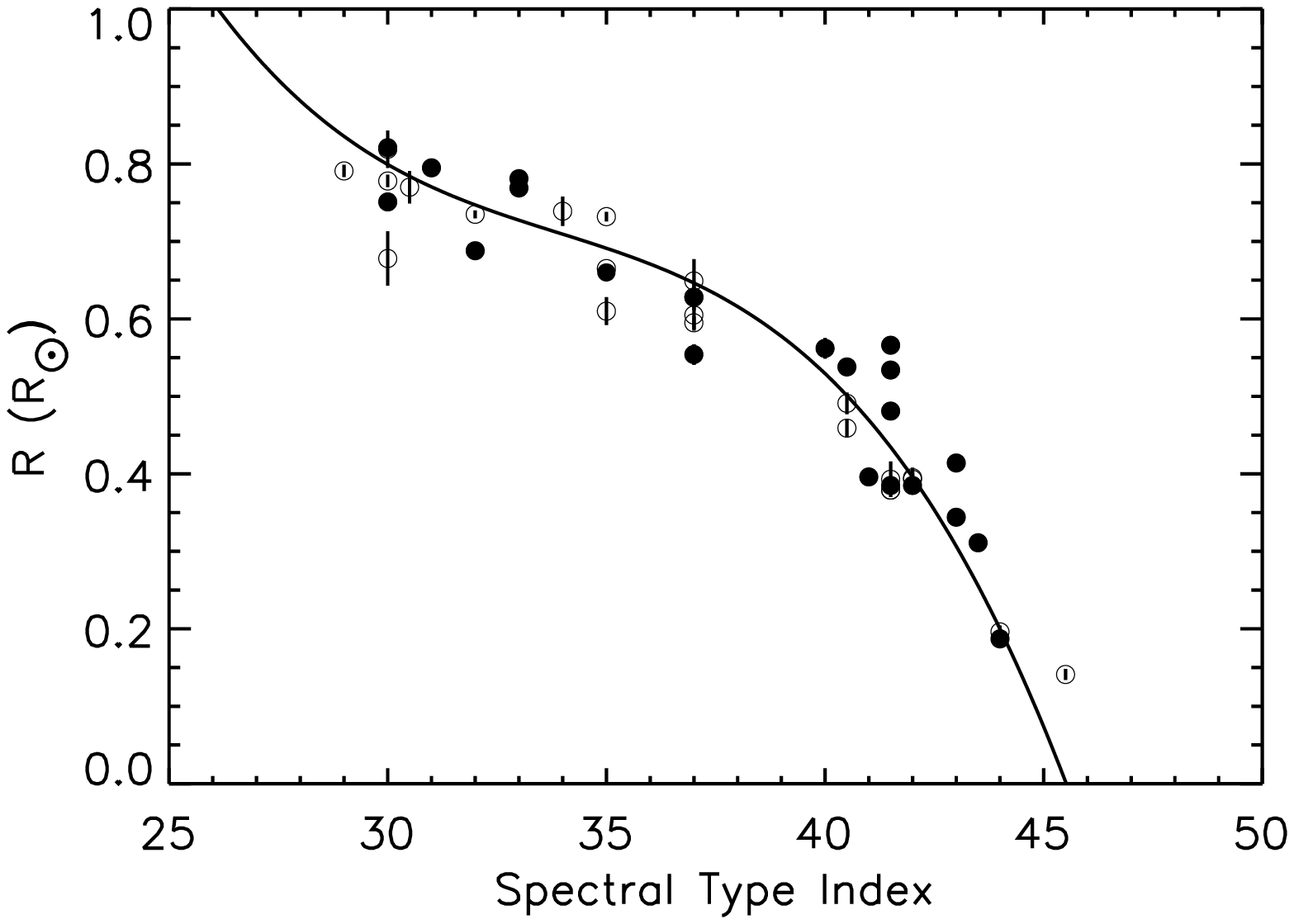}
\caption{Spectral type index code (see text) versus effective temperature (left) and radius (right). Symbols are the same as in Figure~\ref{fig:teff_VS_VmK_and_BmV}.}
\label{fig:teff_and_radius_VS_SpTy}
\end{figure}

We lastly would like to comment on the relations that can be derived with stellar spectral types.  We code the spectral types using a simple linear scale (starting with A0, A1, A2, A3 $\dots$ K0, K1, K2, K3 $\dots \rightarrow 0, 1, 2, 3 \dots 30, 31, 32, 33 \dots$).  Figure~\ref{fig:teff_and_radius_VS_SpTy} shows the results for the relations with temperature and radius. Surprisingly, we find that there is a much tighter correlation with temperature and spectral type (median deviation in $T_{\rm EFF} \sim 75$~K) compared to the color indices mentioned in the above paragraph.  This is likely an artifact of the poor photometry input in the fits also discussed.  In the fit to stellar radii, we can see there is a substantially large spread in values, which in turn produces a fit that is good to only $\sim$~10\%. It is unlikely that this is due to the evolutionary status of the stars since the universe is not old enough for these stars to have evolved from the main sequence.  Perhaps a better explanation for the spread in radii is a difference in composition and/or activity in the stars themselves.

	\subsection{Theory Versus Observation}	

\begin{figure}[!ht]
\plotone{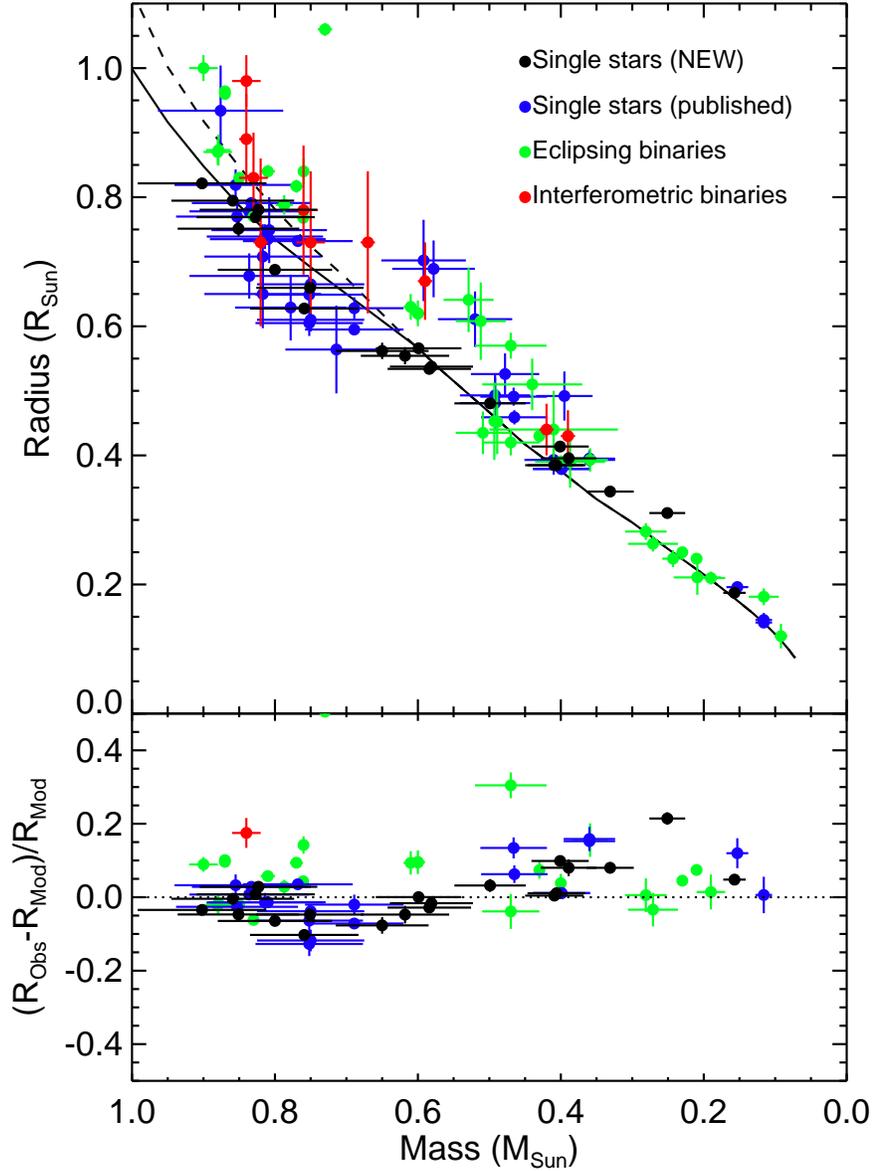}
\caption{{\it TOP:} The Mass-Radius relationship for K and M stars. The solid black line is a 5 Gyr isochrone from the BCAH98 models \citep{bar98}. For stars with mass $> 0.6$~M$_{\odot}$, the dashed line indicates $L_{mix} = H_{p}$ and the solid line  indicates $L_{mix} = 1.9 H_{p}$.  {\it BOTTOM:} Deviation in radius versus mass for stars with radii measurements better than 5\%.  Masses for single stars are derived from the $K$-band mass-luminosity relation from \citet{del00}, and assume a 10\% error.}
\label{fig:mass_VS_radius}
\end{figure}

A reccurring trend seen in the literature is that for stars between $\sim 0.6 - 0.35 M_{\odot}$ (spectral types $\sim$ M0$-$M3.5), theory under-predicts the radius of a star when compared to observations, for both single and binary stars alike (see Figure~\ref{fig:mass_VS_radius}).  For instance, \citet{ber06} used the CHARA Array to measure the radii of six M-type stars, showing that current models are under-estimating their radii by $\sim 15-$20\%, with a possible explanation that some missing source of opacity is not being accounted for and the metallicity of a star is correlated to the deviation of the model predictions to observations.  In fact, \citet{lop07} evaluate this issue in full detail confirming that results are consistent with other published observations of single and binary M-type stars (see \citealt{lop07}, and references therein).  \citet{lop07} find that for the case of binary stars, the activity indicator $L_X/L_{\rm BOL}$ is highly correlated with the radius offset, whereas the single stars (which are several orders of magnitude less active than the binary star sample) have a less strong, but apparent, correlation with the metallicity.  

\begin{figure}[!ht]
\plottwo{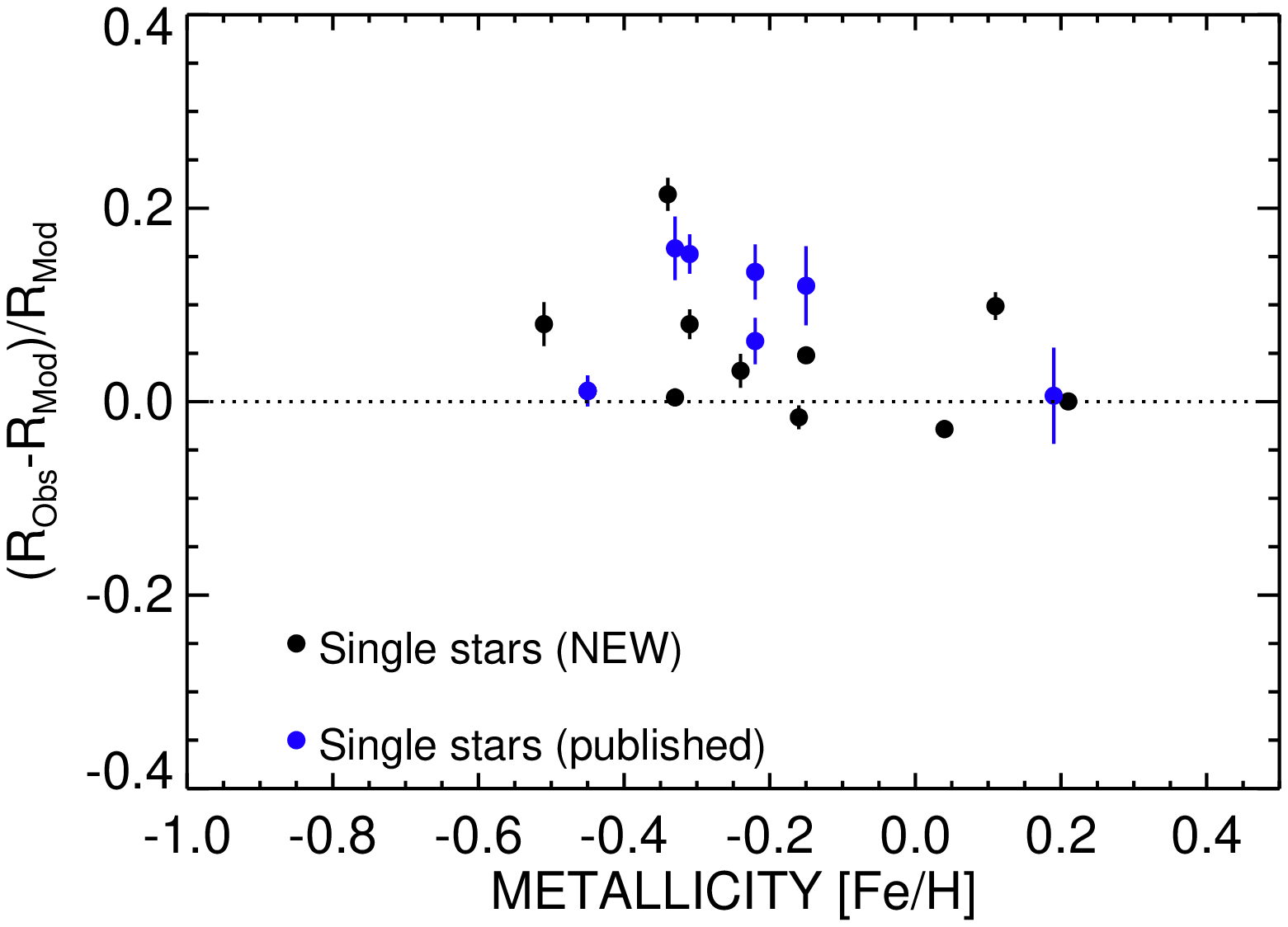}{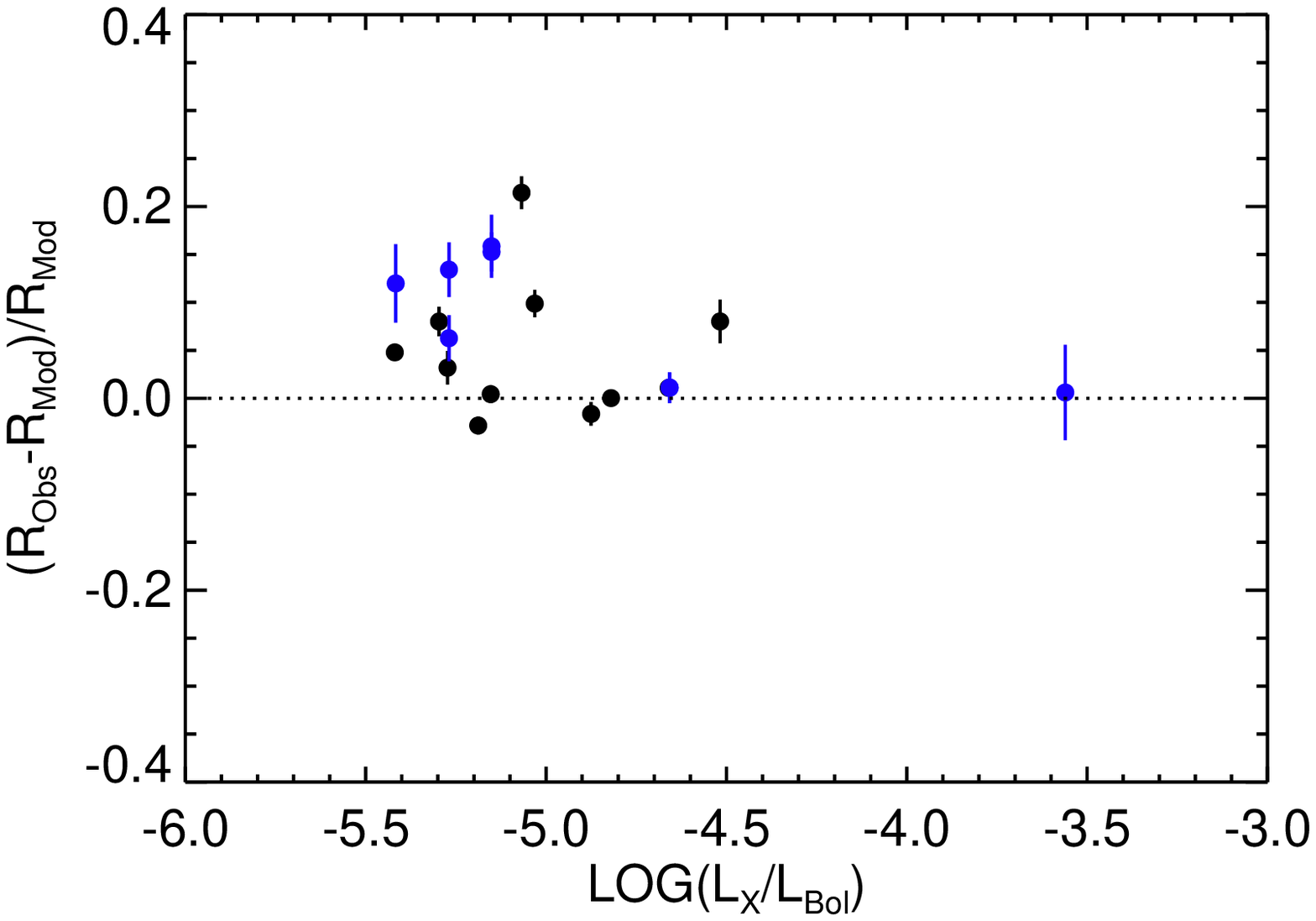}
\caption{Radius deviation of single stars with mass $< 0.6 M_{\odot}$ versus metallicity (left) and $L_X/L_{\rm BOL}$ (right).  The filled black symbols are new results introduced here, whereas the blue symbols are published in the literature.}
\label{fig:radius_offsets}
\end{figure}

With the addition of our new data, which doubles the number of M-type star measurements in this range, these conclusions look a bit different in the context of single stars.  In Figure~\ref{fig:radius_offsets}, we show the fractional radius deviation versus metallicity [Fe/H] and magnetic activity level (expessed as fractional emission of the $X$-ray to the bolometric luminosity of a source $L_X/L_{\rm BOL}$) for single stars with mass $< 0.6 M_{\odot}$.  In these plots, it is apparent that there is still a $\sim 10$\% offset in observed radii versus the model predictions.  However, the introduction of these new data presented here negates the hypothesis that the deviation of the radius prediction has to do with the metallicity and/or activity of the stars.  The offsets are consistent across the full ranges of metallicity and magnetic activity.

We would like to note that this conclusion contradicts \citet{dem09}, who assert that there is no evidence for a radius offset in single stars with respect to the models. To briefly elaborate, their study inspects the entire mass range from $0.9 - 0.1 M_{\odot}$. However, only the early-type M-stars have shown this offset (\citealt{lop07}, and references therein). Furthermore, a closer look at their data set reveals that they only studied new diameters of K-stars (and repeated 2 M-stars).  These choices of data restricted and/or reduced the sensitivity to the detection of an offset.  On the other hand, although we confirm that the disagreement between observations and models still does exist for single stars, the diagnostics explored to shed light on the source of the problem presently reveal nothing to first order.

\section{Summary and Conclusion}

In conclusion, we would like to highlight how a large, homogeneous, and sensitive survey such as this one is crucial to the studies of low-mass stars. Currently, we have measured the diameters of 23 K- and M-type stars, with an average precision of $\sim 1.5$\% using the CHARA Array, a long-baseline optical/infrared interferometer. The magnitude and quality of observations presented for this project doubles the number of low-mass star diameters in the literature, and boosts the precision and sensitivity to the calibration of empirical relations to fundamental properties of stars such as the effective temperature and radius. Our study, with a combination of the highest quality new and extant data, confirms the disagreement between observations and models for single early M-stars at the level of $\sim 10$\%.  However, the diagnostics explored to shed light on the source of the problem are inconclusive. Continuing effort in interferometry and photometry to support precise calibrations is warranted in order to push the limits of our data and extend our knowledge to a large number of stars. The data in hand enable us to further analyze shortfalls in stellar models, where they exist, and provide motivation for adjustments to models in order to match the observations \citep{cha07}.

\acknowledgements TSB and ML-M acknowledges support provided by NASA through Hubble Fellowship grant \#HST-HF-51252.01 awarded by the Space Telescope Science Institute, which is operated by the Association of Universities for Research in Astronomy, Inc., for NASA, under contract NAS 5-26555.  STR acknowledges partial support from NASA grant NNH09AK731. The CHARA Array is funded by the National Science Foundation through NSF grant AST-0908253, by Georgia State University through the College of Arts and Sciences, and by the W. M. Keck Foundation. This research has made use of the SIMBAD literature database, operated at CDS, Strasbourg, France, and of NASA's Astrophysics Data System. This publication makes use of data products from the Two Micron All Sky Survey, which is a joint project of the University of Massachusetts and the Infrared Processing and Analysis Center/California Institute of Technology, funded by the National Aeronautics and Space Administration and the National Science Foundation.

\bibliographystyle{asp2010}            

\bibliography{boyajian_t}

\end{document}